\documentclass[11pt,a4paper]{article}
\pdfoutput=1
\usepackage{jcappub}
\usepackage{bm}

\newcommand{\fnl}{f_{\mathrm{NL}}}

\newcommand{\be}{\begin{equation}}
\newcommand{\ee}{\end{equation}}
\newcommand{\bea}{\begin{eqnarray}}
\newcommand{\eea}{\end{eqnarray}}
\newcommand{\eref}{\eqref}

\newcommand{\kk}{{\bf k}}

\newcommand{\es}{\epsilon_s}

\graphicspath{{./}}
\notoc

\begin{document}

\title{Three-form inflation and non-Gaussianity}

\author[a]{David J. Mulryne}
\author[b,c]{, Johannes Noller}
\author[d]{, Nelson J. Nunes}

\affiliation[a]{Astronomy Unit, School of Physics and Astronomy, Queen Mary University of London, London, E1 4NS, UK}
\affiliation[b]{Theoretical Physics, Blackett Laboratory, Imperial College, London, SW7 2BZ, UK}
\affiliation[c]{Astrophysics, University of Oxford, DWB, Keble Road, Oxford, OX1 3RH, UK}
\affiliation[d]{Faculdade de Ci\^encias e Centro de Astronomia e Astrof\'isica da Universidade de Lisboa, 1749-016 Lisboa, Portugal }

\emailAdd{d.mulryne@qmul.ac.uk}
\emailAdd{johannes.noller@astro.ox.ac.uk}
\emailAdd{njnunes@fc.ul.pt}

\abstract{
We calculate the perturbed action, at second and third order, for a massive three-form field minimally coupled to gravity, and use it to explore the observational predictions of three-form inflation. One intriguing result is that 
the value of the spectral index is nearly independent of the three-form potential, being fixed solely by the number of e-folds 
of inflation, with $n_s=0.97$ for the canonical number of $60$. Considering the bispectrum, we employ  standard techniques 
to give explicit results for two models, one of which produces a large non-Gaussianity. 
Finally, we confirm our results by employing a duality relating the three-form theory to a non-canonical scalar field theory and explicitly 
re-computing results in this dual picture.
}

\keywords{Inflation, non-Gaussianity, bispectrum}

\maketitle
\newpage
        
\section{Background} \label{sec:introduction}

Cosmological inflation \cite{Starobinsky:1980te,Guth:1980zm,Albrecht:1982wi,Hawking:1981fz,Linde:1981mu,Linde:1983gd}, typically assumed to be driven by one or more scalar fields, offers a compelling explanation for the origin of structure in our universe (see \cite{Lidsey:1995np,Lyth:1998xn, Lyth:2009zz,Liddle:2000cg} for reviews). But if inflation is indeed the source of cosmic structure, the field or fields driving this inflationary period have so far not been identified, and a great many candidates have been suggested, including scalar fields with non-canonical kinetic terms \cite{ArmendarizPicon:1999rj, Garriga:1999vw, Silverstein:2003hf}.

During inflation, quantum fluctuations are promoted to classical perturbations, and the statistics of these perturbations are probed by observations (for example \cite{Peiris:2003ff,Komatsu:2010fb}). In recent years, interest in the statistics beyond the two point function, which is parametrised by the power spectrum, has intensified. In particular, numerous investigations of the three-point function, parametrised by the bispectrum, have been undertaken \cite{Maldacena:2002vr, Seery:2005wm, Chen:2006nt}. The bispectrum is potentially a powerful diagnostic, for example it may allow non-canonical single field inflation, which is capable of producing an observable bispectrum \cite{Chen:2006nt}, to be distinguished from canonical inflation, for which the bispectrum is typically too small ever to be observed \cite{Maldacena:2002vr}. 

\subsection{Three-form inflation} 
In addition to scalar fields, other ways of driving inflation have been proposed. These include inflation driven by vector fields (for example \cite{Ford:1989me,Koivisto:2008xf,Golovnev:2008cf}) and higher form fields \cite{Germani:2009iq,KN1,KN2,Germani:2009gg, Koivisto:2009sd,DeFelice:2012jt}. For such models, the techniques needed to make observational predictions are less well explored than for scalar field models. In this paper, our interest lies in the recently proposed inflationary scenarios involving a massive three-form field. Induced potentials for three-forms were studied in the context of string theory in Refs.\cite{Gubser:2000vg,Frey:2002qc}. Inflationary scenarios with three-forms were first investigated by Germani \& Kehagias \cite{Germani:2009iq,Germani:2009gg}, who focused on a non-minimally coupled three-form, with the couplings chosen to make the theory exactly equivalent to a canonical and minimally coupled scalar field theory. 
The behaviour of gravitational waves in this models was studied in Ref.\cite{Kobayashi:2009hj}. 
Later Koivisto \& Nunes \cite{KN1, KN2} considered a minimally coupled massive three-form, which has a novel and much richer associated phenomenology. They found that if 
particular conditions on the potential were satisfied, the model was ghost-free, and moreover could support inflation. More recently, Koivisto \& Urban \cite{Urban:2012ib} studied the non-Gaussian signatures resulting from the magnetic fields generated from the coupling of a three-form to electromagnetism. 

Here we follow Koivisto \& Nunes and consider  the action for a minimally coupled, canonical three-form 
\be \label{3formS}
S = -\int d^4 x \sqrt{-g} \left(\frac{1}{2} R - \frac{1}{48}F^2 - V(A^2) \right)\,,
\ee
where $8\pi G = 1$, $F= \nabla  A$ is a four-form, $A$ a three-form, and we have adopted  a compact notation in which indices are suppressed once the valance of an object is specified,  squaring a quantity denotes contraction of all indices (for example $F^2 \equiv F_{\mu\nu\rho\sigma} F^{\mu\nu\rho\sigma} \equiv \nabla_\mu A_{\nu\rho\sigma} \nabla^\mu A^{\nu\rho\sigma}$), and a dot ($\cdot$) denotes contraction on the first index (for example $\nabla \cdot A=\nabla_\mu A^\mu_{ \rho \sigma}$). Here and throughout Greek indices label space-time dimensions and Roman indices spatial dimensions.
 
In the context of inflationary cosmology, 
the unperturbed three-form must be consistent with the universal symmetries of homogeneity and isotropy. The background three-form can then be written in the form 
\be
A_{ijk}=a(t)^3\epsilon_{ijk}\chi(t)\,
\ee
where $a(t)$ is the scale factor of the universe as a function of cosmic time, and all other elements of $A$ are zero. Hence the unperturbed three-form satisfies $A^2 =6 \chi^2$ \footnote{Note that in Ref. \cite{KN2}, the three form quantity was denoted by $X$ not $\chi$, but here we reserve $X$ for use when we discuss non-canonical scalar field models to match notation commonly used in the literature.}. The dynamics of the universe is then governed by the behaviour of the scalar quantity $\chi(t)$ which is directly related to the three-form. As emphasised elsewhere \cite{KN1,KN2}, when written in terms of this scalar quantity the equations of motion which govern the behaviour of the universe, and the role of the three-form potential, are straightforward to interpret. In particular, adopting units where $8\pi G = 1$, the Friedmann and Raychaudhuri equations are given by 
\begin{eqnarray}
\label{background}
3 H^2 &=&  \frac{1}{2}\left(\dot{\chi} + 3H\chi  \right)^2 + V(\chi^2) \nonumber \\
\dot{H} &=& -\frac{1}{2}V_{,\chi}\chi\,,
\end{eqnarray}
respectively.

On the other hand, as we will review in \S \ref{sec:dualities}, a series of dualities exist which relate the action \eref{3formS} to the action for a massive vector field, and in turn to the action for a massive, and in general non-canonical, scalar field of the form
\be \label{eq:Paction}
S = -\frac{1}{2}\int d^4 x \sqrt{-g} \left ( R - 2 P(X, \phi) \right)\,,
\ee
where $\phi$ is a scalar field and $X=-\nabla^\mu \phi \nabla_\mu \phi$. In principle, therefore, if one wished to understand the dynamics and observational predictions of a particular three-form inflationary model, one could move to working with the dual non-canonical scalar field model, for which suitable techniques are already readily available \cite{Garriga:1999vw,Maldacena:2002vr,Seery:2005wm,Chen:2006nt}. Unfortunately the mapping between theories is rather complicated, and the analytical form of the dual non-canonical scalar field theory is often rather unwieldy or may not even exist, even for fairly simple three-form potentials.

With this in mind, Koivisto \& Nunes \cite{KN1,KN2} advocated working directly with the three-form. Utilising the Einstein field equations, they derived the background equations of motion, and perturbing both the three-form and the metric, the first order perturbation equations. Finally, they formed the Sasaki-Mukhanov equation for the curvature perturbation on comoving hypersurfaces, $\zeta$. This equation follows from the second order perturbed action written purely in terms of $\zeta$, which they thus implied. In addition they were able to address questions such as the stability of the theory, and the  power spectrum produced by three-form inflationary models. Later De Felice, Karwan \& Wongjun \cite{DeFelice:2012jt}  directly perturbed the action, and, following the standard approach, rewrote the result purely in terms of the curvature perturbation. In this way they obtained similar results,  though curiously their perturbed action did not match exactly that of Koivisto \& Nunes. 

\subsection{Overview}

In this paper,  we initially follow Koivisto \& Nunes by working with the three-form theory itself rather than any dual. In contrast to their study, however, we manipulate the action directly, perturbing it to second order and writing the result purely in terms of $\zeta$. Doing so we find agreement with their result. Then we extend existing work by deriving the third order action, which is required to calculate the bispectrum of the curvature perturbation produced during inflation. We also explain why the two previously derived second order actions, by Koivisto \& Nunes and De Felice, Karwan \& Wongjun \cite{DeFelice:2012jt}  respectively, differed slightly. These calculations are covered in \S \ref{sec:actions}. In \S \ref{sec:dualities}, we review the dualities which relate the three-form theory to a non-canonical scalar-field theory. We then show that despite the complexity of the mapping between the theories, we may use this duality to provide an alternative derivation of our second and third order action, providing confirmation of our results. Finally, we employ our results to calculate the bispectrum in two explicit examples of three-form inflation, one of which produces a large non-Gaussianity, in \S \ref{sec:NG}. We conclude in \S \ref{sec:conclusions}.

\section{Perturbations and observables} \label{sec:actions}

In order to make observationally testable predictions for inflation driven by a  three-form field, we must understand the statistics of perturbations produced during the inflationary era. In particular, it is usual to focus on $\zeta$ \cite{Bardeen:1983qw,Lyth:1984gv}, the co-moving curvature perturbation, which is conserved on super-horizon scales if the dynamics is adiabatic \cite{Lyth:2004gb,Rigopoulos:2003ak}.  This property helps  to connect the statistics of the Fourier modes of $\zeta$ soon after they cross the horizon with observations of later epochs.

To this end, we wish to  perturb the action for the three-form theory (see \cite{Mukhanov:1990me,Malik:2008yp,Malik:2008im} for reviews of cosmological perturbation theory). The aim is to determine the form of the autonomous action for $\zeta$ in the setting of three form-inflation. Then by following the standard procedure of quantising $\zeta$ and fixing initial conditions far inside the cosmological horizon, the same-time correlation functions of $\zeta$, produced during inflation, can be calculated using the In-In formalism.   The second order action is required to calculate the two-point correlation function, parametrised by the power spectrum (see for example \cite{Lidsey:1995np} for a review), while the three-point correlation function, parametrised by the bispectrum, requires knowledge of the third order interactions in the theory, and hence the third order action \cite{Maldacena:2002vr, Seery:2005wm, Chen:2006nt}. 

The procedure to calculate the action is to perturb the metric and the three-form, and insert the result into Eq.~\eref{3formS}. Using various constraint equations, it is then possible to write the result purely in terms of only the perturbed quantity $\zeta$, and background quantities. The action can then be separated out order by order. We now present the details of this calculation.

\subsection{The perturbed action for three form inflation}

Following the approach of Maldacena \cite{Maldacena:2002vr}, we consider the Arnowitt-Deser-Misner form of the metric
\be
{\rm d} s^2 = -N^2 {\rm d} t^2 + h_{ij} \left ( {\rm d} x^i + N^i {\rm d} t \right) \left( {\rm d}x^j + N^j {\rm d} t \right),
\ee 
where $N$ is the lapse function, $N_i$ the shift vector, and $h_{ij}$ the spatial metric. In this notation the action for the three-form takes the form
\be
\label{ADMaction}
S=\frac{1}{2}\int {\rm d} t {\rm d} x^3 \sqrt{h} N \left( R^{(3)} - \frac{1}{24} F^2 - 2 V(A^2) \right )+ 
\frac{1}{2} \int {\rm d} t {\rm d} x^3 \sqrt{h} N^{-1} \left( E_{ij} E^{ij} -E^2 \right )\,,
\ee
where $h={\rm det} h_{ij}$, $R^{(3)}$ is the Ricci scalar associated with the spatial metric, 
and $E_{ij} =1/2 \dot{h}_{ij} - N_{(i|j)}$ is the extrinsic curvature. Considering only scalar perturbations, 
when perturbed about an FRW metric the spatial metric takes the form
\be
h_{ij} = a^2 e^{2 \zeta} \delta_{ij}\,,
\ee
where we have used the spatial gauge freedom available to make the spatial metric diagonal, and where $\zeta$ denotes the curvature 
perturbation. 
The lapse function and shift vector are also perturbed to give 
\be
N_i= \psi_{,i} + \tilde{N}_i, ~~~~ N = 1+ \tilde{\alpha}\,,
\ee
where $\tilde{N}_{i,i}=0$ and $\psi$ and $\tilde \alpha$ can be written as an expansion in powers of $\zeta$ as $\tilde \alpha= \tilde \alpha_1 + \tilde \alpha_2$ and $\psi = \psi_1 +\psi_2$. 

We must also account for perturbations in the matter sourcing the Universe's evolution, which in this case is the 
three-form field. The most general form of the perturbed $3$-form is 
\be
A_{0ij} = a(t)\epsilon_{ijk} (\alpha_{,k}+\alpha_k), ~~~~ A_{ijk} = a(t)^3\epsilon_{ijk}(\chi(t) + \alpha_0)\,.
\ee
The vector perturbation $\alpha_k$ together with the vector perturbation in the metric (which we did not write down) decouple from scalar perturbations in 
the second order action and also decay as usual \cite{KN2}. At higher 
order they can also be neglected since our interest here is in the interaction terms which contribute to the scalar-scalar-scalar three-point function. 
The metric tensor perturbation also decouples in the second order action and can be neglected at higher order for the same reason. 
There are two scalar degrees of freedom, $\alpha$ and $\alpha_0$, which just like the metric perturbations can be expanded in powers of $\zeta$.
	
All the scalar perturbed quantities for the both the metric and the matter are then substituted into Eq.~\eref{ADMaction}, which is expanded to third order.

\subsection{Fixing the temporal gauge}
\label{gaugechoice}
Thus far in our calculation we have only fixed the spatial part of the gauge freedom. At first order in perturbation theory, 
under a general gauge transformation, $x^\mu \to x^{u} + \xi^{\mu}$ where $\xi^\mu =  ( \xi^0,\xi_{,i} )$ the scalar three-form 
perturbations transform as 
\be
\alpha_0 \to \alpha_0-\dot{K} \xi^0+X \nabla^2 \xi\,, ~~~~ \alpha \to \alpha - a X \xi\,.
\ee
The freedom associated with $\xi$ is already fixed, but not that associated with $\xi^0$. Considering the transformations above, it is clear that we could fix the gauge by setting $\alpha_0$ to zero, which was the choice made by De Felice, Karwan \& Wongjun \cite{DeFelice:2012jt}. This can also be done at second order.

However, it is important to note that this does not correspond to the choice which is commonly made, 
for example by Maldacena \cite{Maldacena:2002vr}, Seery \& Lidsey \cite{Seery:2005wm} and Chen et al. \cite{Chen:2006nt}, 
who select the comoving gauge, defined at linear order by the condition
\be
\delta T^0_i =0\,.
\ee
In models sourced by a scalar field, this is equivalent to 
setting $\delta \phi$ to zero, while in the present case it implies the following condition
\be
\label{gauge}
\alpha = a^2 \chi \psi \,.
\ee
Enforcing this condition at all orders fixes our temporal gauge freedom, and we can use this relation to replace $\alpha$ with $\psi$, or visa versa. 
We will later  confirm explicitly that this choice is consistent with that made in work on scalar field inflationary models.

Before moving on with our calculation, we note that in their study Koivisto \& Nunes \cite{KN1,KN2} worked in the longitudinal gauge, 
but wrote down the action for the curvature perturbation in the comoving gauge, $\zeta$. The choice of 
De Felice, Karwan \& Wongjun \cite{DeFelice:2012jt} selects a subtly different gauge, and  
explains why the second order actions in these two works do not coincide\footnote{It is important to note, that there is no problem with picking a gauge other than the comoving gauge and then forming $\zeta$ out of quantities in the chosen gauge at the end of the calculation. For example  in Refs \cite{Maldacena:2002vr,Seery:2005gb} the consistency of perturbing the action in the flat gauge and then calculating the statistics of $\zeta$ was shown.}.

\subsection{The second order action}

Moving forward with our calculation, we use the condition \eref{gauge} to substitute for $\alpha$ in the action. We find a rather cumbersome expression containing terms in $\psi$, $\alpha_0$, $\tilde \alpha$ and $\zeta$. The first order part of the action is zero once the background equations \eref{background} of motion are utilised (or equivalently varying the first order part of the action leads to the background equations of motion).  Our aim now is to find substitutions for $\psi$, $\alpha_0$ and $\tilde \alpha$ in terms of $\zeta$, and hence to obtain an action only in terms of $\zeta$. As long as these substitutions follow from a constraint equation, and do not alter the order of the action, we do no harm by using them to write the action in terms of $\zeta$ alone. First we focus on the second order part of the action.

In the present case the simplest way to proceed is to utilise four constraint equations, which are not independent of each other, for the three quantities we wish to eliminate. The first is the equation of motion for $N$, or equivalently $\tilde \alpha$
\begin{eqnarray}
\label{eqmN}
&~&3 H (\dot\zeta-H\tilde\alpha) -\frac{1}{2} (\dot \chi + 3H\chi) (\dot \alpha_0 + 3H\alpha_0) +\frac{1}{2} (\dot\chi+3H\chi)^2 (\tilde \alpha + 3\zeta) \nonumber \\
&~&-\frac{1}{2} V_{,\chi}\alpha_0 + \frac{3}{2} \chi V_{,\chi} \zeta 
+\frac{1}{2} a^2 \chi (\dot \chi + 3 H \chi) \frac{\partial^2}{a^2} \psi - \frac{\partial^2}{a^2} (H\psi+\zeta) = 0.
\end{eqnarray}
The next is the equation of motion for $N_i$, or equivalently $\psi$ 
\be
\tilde \alpha = \frac{\dot \zeta}{H} .    
\label{eqpsi}
\ee
Then we can also use the equation 
which follows from the anti-symmetry of the four-form, and an equation which is most easily seen 
from the conservation of the energy momentum tensor. These are 
\begin{eqnarray}
\label{antisymm}
\alpha_0 &=& 3 \chi \zeta - \frac{V_{,\chi}}{V_{,\chi\chi}} \tilde \alpha, \\
\label{consTmn}
\dot \alpha_0 &=& a^2 \chi \frac{\partial^2}{a^2}\psi + (\dot\chi + 3 H \chi) (\tilde \alpha + 3 \zeta) - 3 H \alpha_0 ,  
\end{eqnarray}
respectively. One can verify that all these equations are self-consistent both with each other and 
with the variation of the action with respect to $\tilde \alpha$, $\psi$ and $\alpha_0$.

We find that the substitutions which consistently satisfy the equations above are 
\begin{eqnarray}
\label{subalpha}
\tilde\alpha &=& \frac{\dot \zeta}{H}, \\ 
\label{subpsi}
\psi &=& -\frac{\zeta}{H} + \sigma, \\ 
\label{subsigma}
\partial^2 \sigma &=& a^2 \frac{\epsilon}{c_s^2} \dot\zeta,
\end{eqnarray}
with $\epsilon = -\dot{H}/H^2 = \chi V_{,\chi}/2 H^2$ and the speed of sound is  given by \cite{KN2}
\begin{equation}
\label{csound}
c_s^2 = \frac{V_{,\chi \chi} \chi}{V_{,\chi}}.
\end{equation}

We note that these substitutions are only first order, while we might expect the second order action to include terms which come from $\tilde \alpha_2$ etc..  As highlighted by Maldacena \cite{Maldacena:2002vr}, however, that these relations come from a constraint equation is advantageous, since it means that only the first order term in the substitutions for the perturbed quantities in terms of $\zeta$ is required. The second order part of each term (for example $\tilde \alpha_2$) will multiply the zeroth order constraint in the second order action. Looking ahead to the third order action, we note that even there we will only need the first order substitution. This is because the second order parts  multiply the first order constraint in the third order action, and so disappear. While the third order part will multiply the zeroth order constraint in the third order action. That this occurs can be verified explicitly.

The next step is to make all these substitutions in the second order part of the action so as to write it purely in terms of $\zeta$. After proceeding in this way, utilising the background equations \eref{background} and  integrating by parts in a suitable manner we arrive at the action
\be
\label{action2a}
S_2= \int {\rm d}t {\rm d}^3x \left [a^3 \frac{\Sigma}{H^2}\dot\zeta^2 - a \epsilon (\partial \zeta)^2 \right ]\,,
\ee
with
\be
\label{sigma}
\Sigma=\frac{H^2 \epsilon}{c_s^2}.
\ee
We note that in this calculation the $\psi$ substitution is not actually used, as $\psi$ naturally cancels out of the second order action after the other substitutions. This second order action matches precisely that derived by Koivisto \& Nunes \cite{KN2}.

\subsection{The third order action}

Now we wish to isolate the third order part of the action in terms of $\zeta$ alone. Once again we can make the substitutions above in the third order part. After numerous integrations by parts we arrive at
\begin{eqnarray}
\label{action3a}
S_3&=&\int \left \{ dt d^3x [-\epsilon a \zeta (\partial \zeta)^2 - a^3 (\Sigma + 2 \lambda) \frac{\dot{\zeta}^3}{H^3}\right.+ \frac{3 a^3 \epsilon}{c_s^2} \zeta \dot{\zeta}^2 \nonumber\\
	&+&\left. \frac{1}{2 a}
\left ( 3 \zeta - \frac{\dot{\zeta}}{H}\right ) \left ( \partial_i \partial_j \psi \partial_i \partial_j\psi - \partial^2 \psi \partial^2 \psi\right) -2 a^{-1} \partial_i\psi\partial_i\zeta \partial^2 \psi\right \}
\end{eqnarray}
where
\be
\label{lambda}
\lambda = -\frac{1}{12} \frac{V_{,\chi}^3 V_{,\chi\chi\chi}}{V_{,\chi\chi}^3}\,.
\ee
and where for simplicity we have not explicitly substituted for $\psi$, but could do so using Eq.~\eref{subpsi}.

At this juncture we pause to make an observation. The form of both the second and third order action, is precisely that found for a $P(X, \phi)$ action of the form \eref{eq:Paction}, except that in that case, $c_s^2$ and $\lambda$ are written in terms of derivatives of $P(X,\phi)$ with respect to $X$ \cite{Seery:2005wm}. One might have expected this, given the formal equivalence of the theories. However, as we have noted previously, the mapping between them is rather complex, and working directly with the scalar dual for a three-form model with a given potential is often cumbersome or indeed impossible, if the analytic form of the inverse potential is not known\footnote{This can be the case, for example, if the potential is a polynomial of degree five or above.}. We will see below, however, that using the duality in a more formal manner allows us to use the $P(X,\phi)$ description to recover Eqs. \eref{csound} and \eref{lambda}, effectively providing an alternative derivation for the perturbed three-form action to third order, and providing a powerful confirmation of our results thus far. We now proceed to that calculation.

\section{Dual theories} \label{sec:dualities}

In this section we review a number of dualities which inter-relate $p$-form theories in four space-time dimensions. These have been discussed elsewhere \cite{Germani:2009iq,Germani:2009gg, Koivisto:2009sd,KN2}, and our aim here is to employ them as a practical tool. To this end we show how one of these dualities can be used to confirm our earlier results for the perturbed action of a three-form theory, utilising previous results derived for a non-canonical scalar field theory by other authors \cite{Seery:2005wm,Chen:2006nt}.

\subsection{Equations of motion for the three-form}

We begin by considering a Palatini-type action, which treats the three-form $A$ and the four-form $F$ as independent variables, and which  is equivalent to \eqref{3formS} up to boundary terms. The Lagrangian for the matter part of the action is given by
\be \label{parent}
{\cal L}_{1} = \frac{1}{48}F^2 - \frac{1}{6}A \nabla \cdot F - V(A^2)\,.
\ee
Variations of the associated action relate $A$ and $F$.  The resulting equations of motion are
\bea 
F &=& -4 \left[ \nabla A \right] \nonumber \\
\nabla \cdot F &=& -12 A V'\left(A^2\right), \label{eom1}
\eea
where a prime denotes differentiation with respect to the argument in brackets. Integrating the middle term of ${\cal L}_1$ by parts inside the action, one finds 
\be \label{ibp}
{\cal L}_{2} = \frac{1}{48}F^2 + \frac{1}{6}F \left[ \nabla A \right] - V(A^2),
\ee
which consequently shares the same equations of motion. The first equation of  \eref{eom1} now appears as a constraint equation and may be substituted back into ${\cal L}_{2}$ to confirm that we do indeed recover~\eqref{3formS}. We note that we are always free to perform such an integration by parts and this procedure leaves the dynamics of the theory invariant. Moreover, we are free to substitute constraint equations  which do not change the order back into the action.

\subsection{Dual actions}

Our primary aim is to rewrite Lagrangians~\eqref{parent} and~\eqref{ibp} in terms of the Hodge ($*$) dual fields to $A$ and $F$. In appendix \ref{appendix1}, for clarity, we give some pedagogical detail regarding such duals. We recall that any $p$-form has a dual ($d-p$)-form, where $d$ is the number of space-time dimensions, four in our context. In particular, the three-form $A$ and four-form $F$ that make up~\eqref{ibp} can be expressed in terms of their duals as 
\begin{align}
\label{hodge}
&(\star F) = \frac{1}{4!}\epsilon_{\alpha\beta\gamma\delta}F^{\alpha\beta\gamma\delta} \equiv \Phi   &F_{\alpha\beta\gamma\delta} = - \epsilon_{\alpha\beta\gamma\delta} \Phi \nonumber\\
&(\star A)_{\alpha} = \frac{1}{3!} \epsilon_{\alpha\beta\gamma\delta} A^{\beta\gamma\delta} \equiv B_\alpha &A_{\beta\gamma\delta} = - \epsilon_{\alpha\beta\gamma\delta} B^\alpha 
\end{align}

The Hodge duals to $F$ and $A$ enable us to recast the original theory \eqref{3formS} into a scalar-vector description, with Lagrangian
\bea \label{scalarvector}
{\cal L}_{3} &=& -\frac{1}{2}\Phi^2 - \Phi \nabla \cdot B - V\left(- 6 B^2\right), 
\eea 
which follows from Lagrangian~\eqref{ibp}.  The equations of motion for $\Phi$ and $B_\mu$ can now be obtained either by varying~\eqref{scalarvector} or equivalently by substituting Hodge duals into~\eqref{eom1}. They are
\bea 
\Phi &=& - \nabla \cdot B , \nonumber \\
\nabla \Phi &=& - 12 B V'(-6B^2).
\label{eq:eomsBPhi}
\eea
The first equation of motion now appears as a constraint equation with respect to ${\cal L}_{3}$ and can be substituted  back to express our original theory~\eqref{parent} as a pure vector theory. Integrating the middle term of ${\cal L}_{3}$ by parts (or equivalently substituting $A$ and $F$ for their duals in ${\cal L}_{2} $), however, one finds that the converse is true, and the second equation of motion in \eref{eq:eomsBPhi} appears as a constraint and may be substituted into the action to remove the vector field in favour of the scalar (or equivalently its dual four form). With an eye on calculating inflationary observables, and in particular the 3-point correlation function, the scalar picture is particularly intriguing, since, as we have discussed, it opens up the possibility of using existing machinery for  dealing with scalar field models of inflation. The final set of equivalent actions for a four-form, three-form, vector and scalar respectively are 
\bea \label{dualtheories}
{\cal L}_{IV}(F,\nabla \cdot F) &=& -\frac{1}{48}F^2 + 2 A^2(\nabla \cdot F) V' \left(A^2(\nabla \cdot F)\right) - V(A^2(\nabla \cdot F)), \\
{\cal L}_{III}(A, \nabla A)) &=& -\frac{1}{3}\left[ \nabla A \right]^2 - V(A^2), \\
{\cal L}_{I}(B,\nabla \cdot B) &=& \frac{1}{2}(\nabla \cdot B)^2  - V\left(-6 B^2\right), \\
{\cal L}_{0}(\Phi, \nabla \Phi) &=& -\frac{1}{2}\Phi^2 -12 B^2(\nabla \Phi) V' \left(-6B^2(\nabla \Phi) \right) - V\left(-6 B^2(\nabla \Phi)\right).
\eea
Here $V\left(-6 B^2(\nabla \Phi)\right)$ and $V(A^2(\nabla \cdot F))$ indicate that the second equation of motion in \eref{eq:eomsBPhi} has to be used in order to express $A^2$ in terms of $\nabla \cdot F$ and $B^2$ in terms of $\nabla \Phi$. Note that we cannot simultaneously use the first equation of motion ($\Phi = - \nabla \cdot B $) to substitute for $-\Phi \nabla \cdot B$ in the action in order to replace it with $\Phi^2$, since this would change the order of the action.

It is interesting to examine the form of these dual theories. The potential for $A/B$ essentially gets mapped into a non-canonical kinetic term for the $F/\Phi$  (4-form/scalar) theory respectively. The canonical kinetic terms in the $A/B$ picture, on the other hand, give rise to simple quadratic potential terms in the corresponding 4-form/scalar theories. This is important for several reasons. First, in this way an effective non-canonical scalar theory arises from a very simple three-form theory.\footnote{The reader might have noticed that there are differences in factors and signs when comparing~\eqref{ibp} and~\eqref{dualtheories} with corresponding equations presented in~\cite{KN2}. The version presented here corrects a small number of typographical mistakes in that work.} 
Second, this immediately tells us that
all scalar models dual to the three-form share
the same simple quadratic potential. This is particularly important for standard
slow-roll inflation\footnote{By this we here mean $\epsilon,\eta... \ll 1$ as well as requiring no rapidly varying speed of sound.}, where in the dual scalar picture the potential dominates over the kinetic terms. The
fact that all
models share the same dual scalar potential then implies that the form of the original
three-form potential (which turns into a non-canonical kinetic term) is not important
when computing, for example, the spectral
index $n_s$. We will return to this point later.

Finally, one may wonder how one can come up with an effective single scalar theory dual to a three-form theory, which in principle possesses more physical degrees of freedom. However, starting with the most general canonical and minimally coupled 3-form action in 4$d$, as we do here, guarantees that such a dual single scalar field description always exists. This is the case, because 1) the canonical kinetic term for the three-form dualises to a $\Phi^2$ potential 2) the three-form potential is a function of $A^2$ only, because in 4d this is the only covariant scalar combination that can be built from a 3-form - thus the potential only depends on one effective degree of freedom: $A^2$ and 3) these two degrees of freedom, $\Phi$ and $A^2$, are related via an equation of motion, leaving only one effective independent degree of freedom, thus explaining the existence of a dual single scalar description.

\subsection{The perturbed action from scalar duality}
 
Now we wish to show that one can derive Eqs. \eref{csound} and \eref{lambda} using the results above. In particular, we employ the expressions 
\be
B^2 = -\chi^2, \hspace{1cm} X \equiv -g^{\mu\nu} \nabla_\mu \phi \nabla_\nu \phi = -12^2 B^2 \left(V'(-6B^2)\right)^2 \,,
\ee
which follow from the definition of the Hodge dual to $A$, Eq.~\eref{hodge}, and from Eq.~\eref{eq:eomsBPhi} respectively, to find  
\be
\label{derv}
X = V_{,\chi}^2, \hspace{1cm} {\rm and} \hspace{1cm} 
\frac{\partial \chi}{\partial X} = \frac{1}{2 V_{,\chi} V_{,\chi \chi}}\,.
\ee 
From the discussion of dualities, and in particular ${\cal L}_{0}$, we see that the $P(X,\phi)$ theory dual to our minimally coupled three-form theory was parametrised by
\begin{eqnarray}
P(X,\phi) &=& - \frac{1}{2} \phi^2 -12B^2 V'(-6B^2) -V(-6B^2)\nonumber \\
&=&  -\frac{1}{2} \phi^2 + \chi V_{,\chi} - V(6\chi^2).
\end{eqnarray}
Differentiating this expression with respect to $X$ and using \eref{derv} we find $P_{,X}$, $P_{,XX}$ and $P_{,XXX}$ which upon substitution into
\be
\label{csTF}
c_s^2 = \frac{P_{, X}}{P_{, X}+2  X P_{,X X}}, \hspace{1cm} \lambda = X^2 P_{,XX} +\frac{2}{3} X^3 P_{,XXX}\,,
\ee
which are the expressions for $c_s^2$ and $\lambda$ for a $P(X,\phi)$ theory \cite{Seery:2005wm}, confirms the expression for the speed of 
sound \eref{csound} and for $\lambda$ \eref{lambda} we found for the three-form theory.

This constitutes the result we are looking for. If we had only known about the perturbed action in terms of $\zeta$ from a $P(X,\phi)$ theory, we could have used the argument above to write that perturbed action   solely in terms of three-form quantities, hence providing an alternative route to our earlier result.

The power of this approach is that we can also readily probe higher order statistics, such as the trispectrum, since the quartic action has already been calculated for $P(X,\phi)$ theories \cite{Huang:2006eha,Arroja:2008ga}. This involves the quantity $\Pi$, defined in Ref.~\cite{Arroja:2008ga}, which for completeness, we calculate to be
\be
\Pi = \frac{1}{40} \frac{V_{,\chi}^3 V_{,\chi\chi\chi}}{V_{,\chi\chi}^3}
+ \frac{3}{40} \frac{V_{,\chi}^4 V_{,\chi\chi\chi}^2}{V_{,\chi\chi}^5}
- \frac{1}{40} \frac{V_{,\chi}^4 V_{,\chi\chi\chi\chi}}{V_{,\chi\chi}^4},
\ee
in terms of the three-form quantity $\chi$.

\subsection{Confirmation of our gauge choice}
Finally we would like to confirm that the gauge we have chosen in \S \ref{gaugechoice} is equivalent to the gauge chosen when the perturbed 
action for the $P(X,\phi)$ theory is calculated \cite{Seery:2005wm, Chen:2006nt}. There the choice is usually expressed as $\delta \phi = 0$. Utilising the expression 
\be
X=12 \chi^2 \left(V'(6\chi^2)\right)^2\,,
\ee
and perturbing both sides, we find this leads to the expression
\be
\alpha_0 = 3 \chi \zeta - \frac{V_{,\chi}}{V_{,\chi\chi}} \tilde\alpha,
\ee
which is the same as the anti-symmetry condition \eref{antisymm} after we fixed the gauge choice $\alpha = a^2 \chi \psi$. It is then possible to verify that the symmetry constraint on the three-form gives Eq. \eref{gauge}, and the other substitutions remain unchanged. In this way the equivalence of our gauge choice with the usual choice made in scalar field theories is confirmed, in contrast with the choice made in other works \cite{DeFelice:2012jt}.

\section{Non-Gaussianities}
\label{sec:NG}
\subsection{Correlation functions}

We will now proceed to utilise our results thus far to compute  observables for the three-form inflationary theory under consideration. Because our action is identical to that for a non-canonical scalar field (except with $c_s^2$ and $\lambda$ being expressed in terms of background three-form quantities), the calculation follows precisely that of Garriga \& Mukhanov \cite{Garriga:1999vw} to calculate the power spectrum, and  Seery \& Lidsey \cite{Seery:2005wm} and Chen et al. \cite{Chen:2006nt} to calculate the three-point function. Here we simply provide an overview of the main steps and the important results.

The 2-point correlation function for the curvature perturbation $\zeta$ is defined as
\begin{equation} \label{2point}
\langle \zeta(\textbf{k}_1)\zeta(\textbf{k}_2)\rangle = (2\pi)^5
\delta^3(\kk_1+\kk_2)  \frac{P_\zeta}{2 k_1^3}\,,
\end{equation}
and is calculated using the second order action Eq.~\eref{action2a}. As found by \cite{Garriga:1999vw} the end result is    
\begin{equation}
P_\zeta \equiv \frac{1}{2\pi^2}k^3\left\vert\zeta_k\right\vert^2 =\left. \frac{1}{2 (2 \pi)^2\epsilon}\frac{{ H}^2}{{ c}_sM_{\rm Pl}^2}\right|_*\,,
\label{zetaPk}
\end{equation}
where $*$ indicates that the expression is evaluated at horizon crossing $c_s k = a H$. The spectral index $n_s$ is then given by
\be \label{ns}
1-n_s = 2\epsilon +\frac{\dot \epsilon}{\epsilon H} +\frac{\dot c_s}{c_s H}\,.
\ee

The slow-roll approximation (i.e. neglecting $\ddot\chi$ in the equation of motion for $\chi$) implies that 
\be 
\label{chiN}
\frac{\chi_{,N}}{\chi} \approx -\frac{2}{3}\frac{1}{\chi^2}\left(1-\frac{3}{2}\chi^2\right)\epsilon .
\ee
Substituting this into the Friedmann equation we obtain \cite{KN1}
\be
\label{epsiloneq}
\epsilon \equiv - \frac{\dot H}{H^2} \approx \frac{3}{2} \chi \frac{V_{,\chi}}{V} \left(1- \frac{3}{2}\chi^2\right) .
\ee
Eliminating the term in brackets from the last two equations we can write 
\be 
\frac{\chi_{,N}}{\chi} = -\frac{4}{9}\frac{1}{\chi^2} \frac{V}{\chi V_{,\chi}} \epsilon^2 .
\ee
This expression allows us to easily compute $\dot \epsilon$ from the definition of $\epsilon$ to find $\dot\epsilon/\epsilon H = 2\epsilon + {\cal O}(\epsilon^2)$. Similarly, we find that $\dot c_s/c_s H \approx 
{\cal O}(\epsilon^2)$ 
and also $\dot \lambda/\lambda H \approx {\cal O}(\epsilon^2)$.
This means that to first order in the slow-roll parameters, the scalar spectral index is simply $n_s = 1 - 4\epsilon$.

The ratio of tensor to scalar perturbations was computed in Ref.\cite{KN2} and shown to be related to the slow roll parameter $\epsilon$ as\footnote{A more exact answer, fully taking into account that tensor and scalar modes freeze out at different times for models with $c_s \ne 1$, can be found in \cite{Agarwal:2008ah}. Here we just note that this extra effect means that the actual tensor-to-scalar ratio $r$ is in fact smaller than naively expected from $r = 16 c_s \epsilon$ for $c_s < 1$.}
\be
\label{eqr}
r = 16 c_s \epsilon,
\ee
mirroring the analogous expression for non-canonical scalar field models. One can also compute the spectral index for tensor perturbations $n_t$, finding $n_t = -2\epsilon$ to first order in slow-roll parameters as usual.

The third order action \eref{action3a} is needed to calculate the three-point correlation function. 
First, however, it must be written in a suitable form, which follows from the use of the $\psi$ substitution, and is \cite{Seery:2005wm,Chen:2006nt}
\begin{eqnarray} \label{action3b}
S_3&=&\int {\rm d}t {\rm d}^3x \left\{
-a^3 \left[\Sigma\left(1-\frac{1}{c_s^2}\right)+2\lambda\right] \frac{\dot{\zeta}^3}{H^3}
+\frac{a^3\epsilon}{c_s^4}(\epsilon-3+3c_s^2)\zeta\dot{\zeta}^2 \right.
\nonumber \\ &+&
\frac{a\epsilon}{c_s^2}(\epsilon-2\es+1-c_s^2)\zeta(\partial\zeta)^2-
2a \frac{\epsilon}{c_s^2}\dot{\zeta}(\partial
\zeta)(\partial \sigma) \nonumber \\ &+& \left.
\frac{a^3\epsilon}{2c_s^2}\frac{d}{dt}\left(\frac{\eta}{c_s^2}\right)\zeta^2\dot{\zeta}
+\frac{\epsilon}{2a}(\partial\zeta)(\partial
\sigma) \partial^2 \sigma+\frac{\epsilon}{4a}(\partial^2\zeta)(\partial
\sigma)^2+ 2 f(\zeta)\left.\frac{\delta L}{\delta \zeta}\right\vert_1 \right\},
\end{eqnarray}
where $f(\zeta) \delta L/\delta \zeta|_1$ indicates terms proportional to the functional derivative of the Lagrangian evaluated at first order in $\zeta$.  
This would be zero if $\zeta$ was Gaussian. Such terms can be removed by a field redefinition (though one 
must recall the redefinition when forming correlation functions so as to form corrections of $\zeta$ itself). 
This form of the action thus identifies the relevant interaction vertices which contribute towards the 
three-point function.

At tree level in quantum field theory, and in the interaction picture, the In-In (equal time) 
three-point correlation function is given by the expression
\begin{equation} \label{interaction}
\langle
\zeta(t,\textbf{k}_1)\zeta(t,\textbf{k}_2)\zeta(t,\textbf{k}_3)\rangle=
-i\int_{t_0}^{t}{\rm d}t^{\prime}\langle[
\zeta(t,\textbf{k}_1)\zeta(t,\textbf{k}_2)\zeta(t,\textbf{k}_3),H_{\rm int}(t^{\prime})]\rangle ~,
\end{equation}
where $H_{\rm int}$ is the Hamiltonian evaluated at third order in the perturbations and follows directly 
from~\eqref{action3b}. Vacuum expectation values are evaluated with respect to the interacting vacuum $|\Omega \rangle$. 
By convention the 3-point correlation function is parametrised by the amplitude ${\cal A}$.
\begin{equation} \label{amplidef}
\langle \zeta(\textbf{k}_1)\zeta(\textbf{k}_2)\zeta(\textbf{k}_3)\rangle = (2\pi)^7
\delta^3(\kk_1+\kk_2+\kk_3) P_\zeta^{\;2} \frac{1}{\Pi_j k_j^3}{\cal A}\,,
\end{equation}
where, again by convention the power spectrum $P_\zeta$ in the above formula is calculated for the mode 
$K = k_1 + k_2 + k_3$. Evaluating \eref{interaction}, one can determine ${\cal A}$

In principle ${\cal A}$ is a general function of the three Fourier modes, which are related by the 
condition $\textbf{k}_1 + \textbf{k}_2 + \textbf{k}_3=0$ (within the slow-roll approximation the full form of ${\cal A}$ is given in Ref.~\cite{Chen:2006nt}). 
For a given shape of non-Gaussianity, however, (see figure~\ref{fig1}) the size of non-Gaussianity can 
be adequately characterised by a single-value measure $f_{\rm NL}$. For an equilateral shape (i.e. one peaking in the limit $k_1 \sim k_2 \sim k_3$), this can be defined as~\cite{Khoury:2008wj}
\be
f^{\rm equil}_{\rm NL} = 30\frac{{\cal A}_{k_1=k_2=k_3}}{K^3}\,,\label{fnl}
\ee
where amplitudes are matched at $k_1=k_2=k_3=K/3$. Note that we follow the WMAP sign convention here, where positive $f_{\rm NL}$ physically corresponds to negative-skewness for the temperature fluctuations. The parameters controlling the overall size of $f_{\rm NL}^{\rm equil}$ are $c_s,\Sigma,\lambda$, which we now summarise for convenience:
\begin{align} \label{3formCSL}
&c_s^2 = \frac{V_{,\chi\chi} \chi}{V_{,\chi}},&
&\Sigma = \frac{V_{,\chi}^2}{2 V_{,\chi\chi}},& 
&\lambda = - \frac{V_{,\chi}^3 V_{,\chi\chi\chi}}{12 V_{,\chi\chi}^3}.&
\end{align}  
Following~\cite{Chen:2006nt} we may now compute $f^{\rm equil}_{\rm NL}$ in the slow-roll regime, finding the result at leading order to be\footnote{Note that \eqref{eqfnl} differs from the expression found by Chen et al.~\cite{Chen:2006nt} by an overall sign. This is because Chen et al. use a sign convention opposite to that used by WMAP and throughout this paper.} 
\be 
\label{eqfnl} 
\fnl^{\rm equil} \approx \frac{5}{81}\left(\frac{1}{c_s^2}-1- 2 \frac{\lambda}{\Sigma}\right) 
- \frac{35}{108} \left(\frac{1}{c_s^2}-1\right) +
 {\cal O}\left(\epsilon,\frac{\epsilon}{c_s^2}, \epsilon \frac{\lambda}{\Sigma},\frac{\dot \lambda}{\lambda H}\right).
\ee

This expression clearly shows that models with small speed of sound can lead to a large non-Gaussian signal. In the context of three-forms, a hyperbolic potential was discussed in Ref. \cite{DeFelice:2012jt}, which yields a small speed of sound during inflation, though close to unity in the oscillatory phase. Here we explore two other simple potentials.

\subsection{Example I: A power-law potential} \label{subsec-power}

We now consider some concrete examples, and begin by considering a three-form model self-interacting through 
a simple power-law potential
\be \label{powerlawL}
{\cal L} = -\frac{1}{48}F^2  - V_0 A^{2p}\,,
\ee
that is, $V(A^2) = V_0 A^{2p} = V_0 \left(6 \chi^2\right)^p$, where $p$ is a constant. 

This is a special example, where the equivalent $P(X,\phi)$ theory is relatively simple, and is given by
\be
{\cal L}_{\phi} = (2p - 1) \left(\frac{1}{V_0}\right)^{1/(2p-1)}\left( \frac{X}{24 p^2}\right)^{\frac{p}{2p-1}} - \frac{1}{2}\phi^2,
\ee
where one now sees explicitly that the 3-form potential has been mapped into a non-canonical kinetic term for the effective scalar $\phi$.
As long as $2p - 1 \ne 0$, one can check that equations~\eqref{csound}, \eqref{sigma} and~\eqref{csTF} equivalently yield
\begin{eqnarray}
c_s^2 &=& 2p - 1 \,,\nonumber \\
\frac{\lambda}{\Sigma} &=&  \frac{1}{3}\frac{1 - p}{2p-1}\,, \label{csls}
\end{eqnarray}
and we can immediately make observational predictions for the theory, finding several interesting results.

Here both the speed of sound and $\lambda/\Sigma$ are constant, and we have $n_s -1 = -4\epsilon$. The constancy of the speed of sound has the interesting consequence that the observational requirement of obtaining a (near)-scale-invariant spectral index $n_s$ forces $\epsilon$ to be close to zero and slowly-varying here (compare \eqref{ns} and also \cite{Khoury:2008wj}). For this power-law model, slow-roll therefore becomes an observational requirement in contrast to models with varying speed of sound.

\begin{figure}[tp] 
\begin{center}$
\begin{array}{c}
\includegraphics[width=0.7\linewidth]{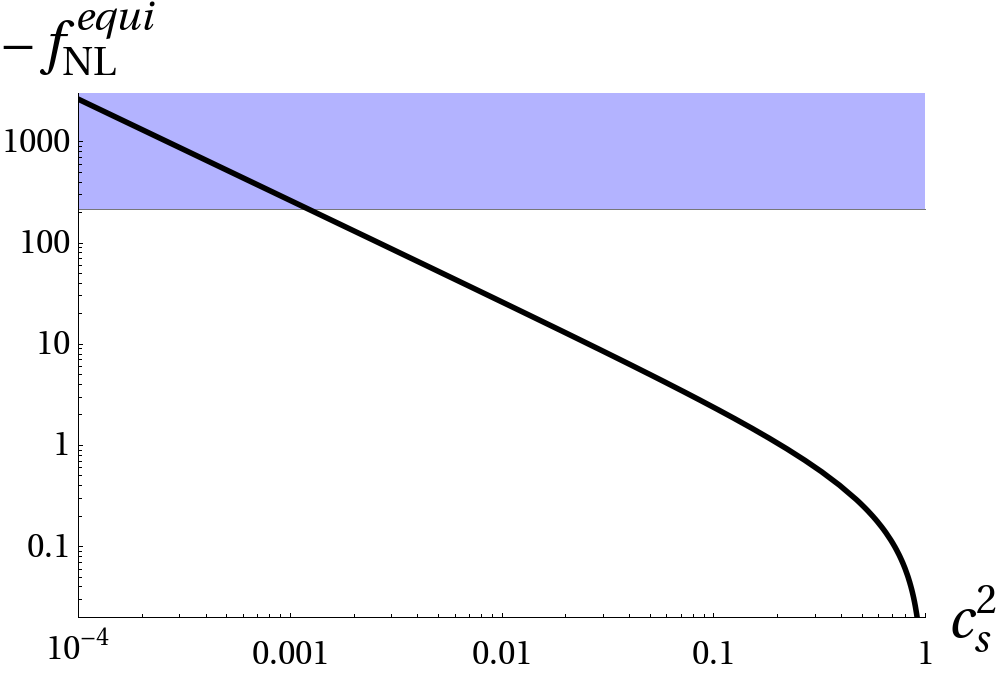} 
\end{array}$
\end{center}
\caption{
Dependence of $f_{\rm NL}^{equil}$ on ${c}_s^2$ for the power law potential $V = V_0 (6\chi^2)^p$ and $N = 60$. A large and generically negative non-Gaussian amplitude is found. The shaded region is disallowed by the WMAP 2$\sigma$ bound $-214 < f_{\rm NL}^{equil} < 266$.\cite{Komatsu:2010fb} We recall that $c_s^2 = 2p - 1$ for this model. 
\label{tejo1fnl}}
\end{figure}

\begin{figure}[tp]
\begin{center}
\includegraphics[width=0.7\linewidth]{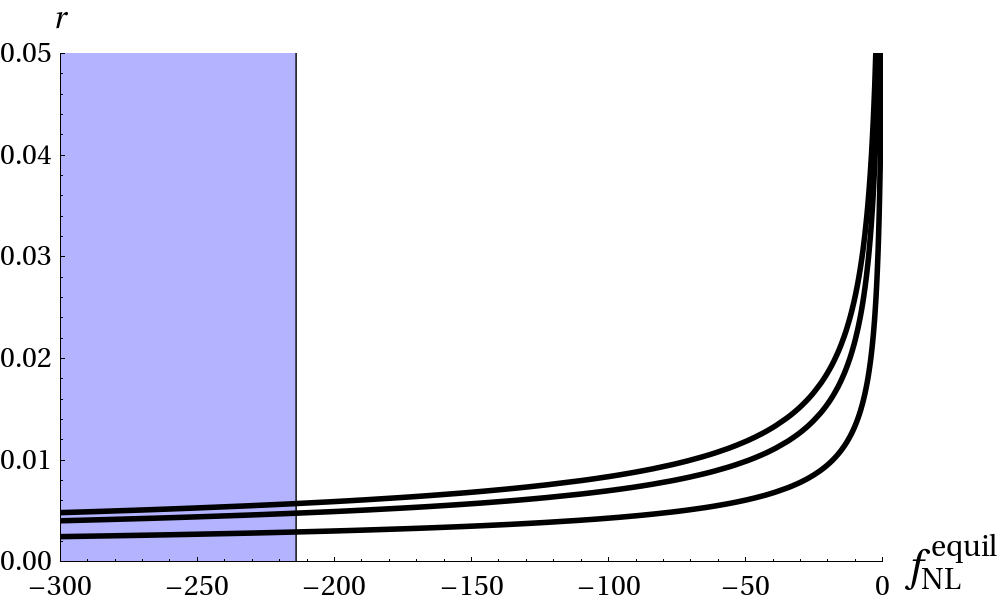}
\caption{\label{fnlr} The solid lines show how the parameters $f_{NL}^{\rm equil}$ and $r$ are related to each other for the power law potential, $V = V_0 (6\chi^2)^{p}$ and for $N = 50, 60, 80$ from top to bottom. The shaded region is disallowed by the WMAP 2$\sigma$ bound $-214 < f_{\rm NL}^{equil} < 266$.\cite{Komatsu:2010fb}}
\end{center}
\end{figure}

We now need to calculate this quantity $N$ $e$-folds before the end of inflation. 
It was shown in \cite{KN1} that for a power law potential, the value of the field at this time can be estimated to be
\be
\chi_N^2 = \frac{2}{3} - \frac{4}{18 p} \frac{1}{1+2N},
\ee 
which upon substitution in \eqref{epsiloneq} gives
\be
\epsilon_N \approx \frac{1}{1+2N}.
\ee
Assuming that $N\approx 60$ is required to solve the horizon problem, we predict that the spectral index on observationally 
relevant scales for a three-form with power law potential is
\be
n_s \approx 0.97,
\ee
and independent of the value of the power $p$. Naturally the spectral index will be closer to scale-invariant the longer inflation lasts. 
Substituting for $c_s^2$ and $\lambda/\Sigma$ into \eref{eqfnl} we obtain the dependence of $f_{NL}^{\rm equil}$ on $c_s^2$ 
illustrated in Fig.~\ref{tejo1fnl}. It clearly  shows that a small speed of sound leads to a large $f_{NL}^{\rm equil}$ as expected.
Substituting for the speed of sound for a power law potential in Eqs.\eref{eqr} and \eref{eqfnl} we can relate $f_{NL}^{\rm equil}$ and $r$. Figure~\ref{fnlr} illustrates how these two quantities are related, and shows the region of the parameter space ($r,f_{NL}$) that is allowed given current bounds. 

\begin{figure}[tp] 
\begin{center}$
\begin{array}{c}
\includegraphics[width=0.7\linewidth]{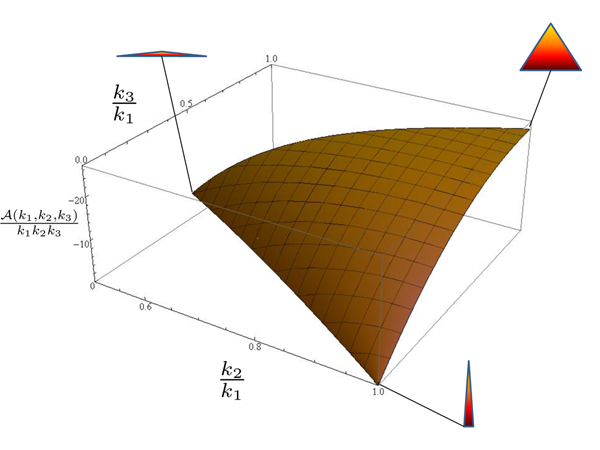} \\
\end{array}$
\end{center}
\caption{
The dimensionless bispectrum ${\cal A}(k_1,k_2,k_3)/k_1 k_2 k_3$ for the power law potential, $V = V_0 (6\chi^2)^{p}$ with $p = 0.505$, corresponding to $c_s = 0.1$, and with $N = 60$. Triangular shapes denote the equilateral, squeezed/local and enfolded limit clockwise from top right. A predominantly equilateral shape is found. 
\label{fig1}}
\end{figure}

Phenomenologically we find that no large non-Gaussian amplitude of the enfolded or orthogonal types (which have peaks in the folded limit $k_1 \sim k_2 \sim 2 k_3$) can be generated here in contrast to generic single field inflation models. To see why, it is useful to notice that one may express $\lambda/\Sigma$ as \cite{Seery:2005wm,Khoury:2008wj}
\be \label{fX}
\frac{\lambda}{\Sigma} = \frac{1}{6} \left( \frac{2 f_X + 1}{c_s^2} - 1 \right)
\ee
where
\be
f_X = \frac{\epsilon \epsilon_s}{3 \epsilon_X}, \quad\quad\quad \epsilon_s = \frac{\dot{c_s}}{H c_s}, \quad\quad\quad \epsilon_X = -\frac{\dot{X}}{H^2}\frac{\partial H}{\partial X}.
\ee
Now, since the speed of sound is constant for the power-law model, $\epsilon_s, f_X$ are identically zero here. A large and predominantly orthogonal (or enfolded) amplitude, however, requires a negative and non-zero $f_X$ (assuming positive $\epsilon$ and $0 \leq c_s \leq 1$)\cite{Senatore:2009gt}\footnote{In general $P(X,\phi)$ inflation models a predominantly orthogonal or folded shape can be generated by finely balancing the contributions from the $\dot{\zeta}^3$ interaction vertex (which depends on $\lambda/\Sigma$) against the other vertices such that the generically predominant equilateral shape contributions cancel out. In this way the otherwise subdominant orthogonal or folded configurations are brought out. As stated above, this delicate cancellation relies on $f_X$-dependent contributions though, whereas $f_X = 0$ identically for the power-law example given.}. 
As such the non-Gaussian shape found here is always predominantly equilateral. Furthermore, since parameters $\lambda$ and $\Sigma$ are simply related to $c_s$ via \eqref{fX}, the non-Gaussian amplitude is completely controlled by $c_s^2$ for a power-law potential. This means that $f_{\rm NL}^{\rm equil} \sim {\cal O}(c_s^{-2})$ for subluminal and positive speed of sound $c_s$, i.e. $p \; \epsilon \left[1/2,1\right]$. Figure~\ref{fig1} illustrates these points. Finally note that, analogous to e.g. DBI inflation, any sizable level of non-Gaussianity has a negative $f_{\rm NL}^{\rm equil}$ associated with it here.

\subsection{Example II:  Exponential potential}

Let us now consider an alternative potential. Even though it is fairly simple in its analytical form, the exponential potential, $V = V_0 \exp(\beta A^2)  = V_0 \exp(6 \beta \chi^2)$, becomes very complex when written as a $P(X,\phi)$ theory,
\be
{\cal L}_0 = \left(W(x) -1\right) V_0 \exp\left(\frac{1}{2} W(x)\right) - \frac{1}{2}\phi^2,
\ee
where $W(x)$ is the Lambert-W function and $x = X/12\beta V_0^2$. Dealing with this model in the $P(X,\phi)$ description is therefore very difficult. This is a good example showing that performing the calculations in the original three-form theory is far simpler than going to the $P(X,\phi)$ description.

Inflation ends when $\epsilon \approx 1$ which for this potential takes place when
\be
\chi_e^2 = \frac{1}{3} + \frac{1}{3} \left(1-\frac{1}{3\beta}\right)^{1/2} .
\ee
This expression implies that inflation only ends if $\beta >1/3$. In particular, if $\beta \gg 1/3$ we can approximate
\be
\chi_e^2 \approx \frac{2}{3} - \frac{1}{18\beta}.
\ee
In this case, the solution of \eref{chiN} and \eref{epsiloneq} is non-analytical and we must resort to an approximation. Defining $\chi^2 = 2/3 - y$, since we know that inflation occurs very close to $2/3$, we obtain
\be
\chi_N^2 = \frac{2}{3} - \frac{1}{18\beta} \, \frac{1}{1+\sqrt{6} N}.
\ee
The slow-roll parameter $N$ $e$-folds before the end of inflation is then $\epsilon_N \approx 1/(1+\sqrt{6}N)$. Here, $c_s^2 = 1+ 12 \beta \chi^2$ and it can be verified that $\dot{c}_s/c_sH  = -\epsilon^2/\chi^2 \sim {\cal O}(\epsilon^2)$, hence, for $N = 60$, we obtain for this model
\be
n_s \approx 0.97,
\ee
independent of the value of $\beta$. The fact that the value of the scalar spectral index is independent of parameter $\beta$, or $p$ in the case of the power law potential, and we obtain the same value in both examples, should not come as a surprise. The reason becomes clear when considering the 
dual $P(X,\phi)$ theory, which for any three-form potential, has the same quadratic scalar potential. Since we are 
in a slow-roll regime, the functional form of the kinetic term is not important with respect 
to the potential which leads to identical results for the spectral index.

It was found above that the choice of the exponential three-form potential offers an exit from inflation only if $\beta > 1/3$. 
Taking $\chi_N^2 \approx 2/3$, this puts the lower bound, $c_s^2 \gtrsim 11/3$, which means that the speed of sound is superluminous. 

\begin{figure}[tp] 
\begin{center}$
\begin{array}{c}
\includegraphics[width=0.7\linewidth]{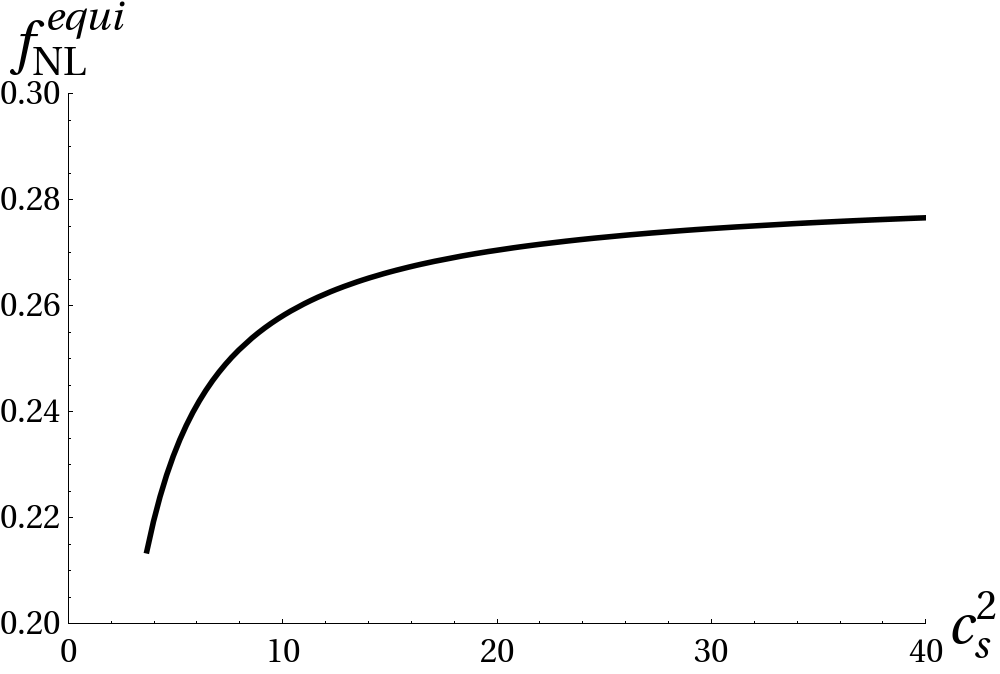} 
\end{array}$
\end{center}
\caption{
Dependence of $f_{\rm NL}^{equil}$ on ${c}_s^2$ for the exponential potential, $V = V_0 \exp(6\beta\chi^2)$ and $N = 60$. A small and generically positive non-Gaussian amplitude is found. Notice the lower bound $c_s^2 \gtrsim 11/3$ for this model.
\label{tejo2fnl}}
\end{figure}

\begin{figure}[tp]
\begin{center}
\includegraphics[width=0.7\linewidth]{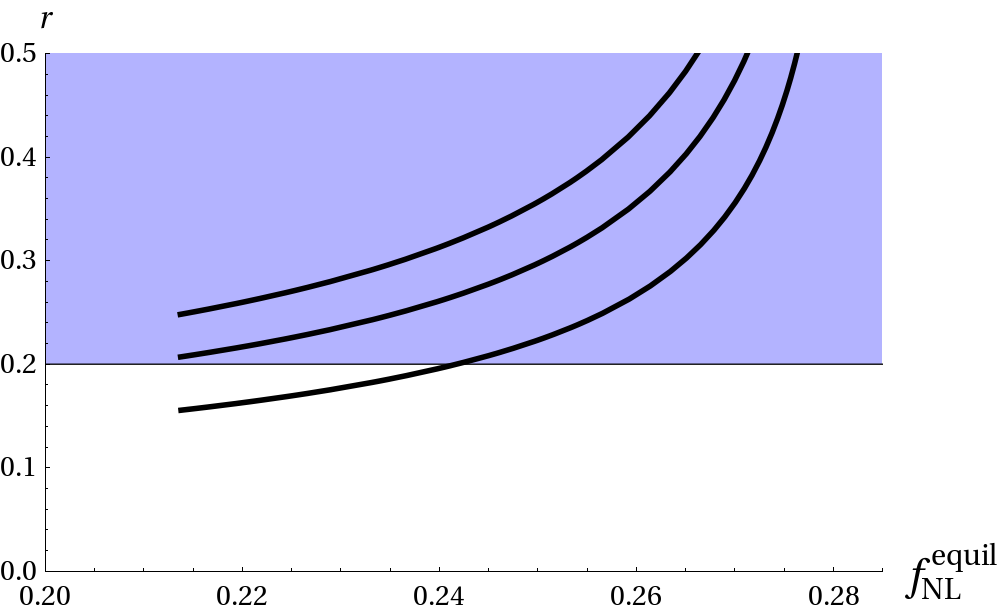}
\caption{\label{tejo2rfnl} The solid lines show how the parameters $f_{NL}^{\rm equil}$ and $r$ are related to each other for the exponential potential, $V = V_0 \exp(6 \beta \chi^2)$ and for $N = 50, 60, 80$ from top to bottom. The shaded region is disallowed by the WMAP 2$\sigma$ bound $r \lesssim 0.2$.\cite{Komatsu:2010fb}. End points for the solid lines at $f_{\rm NL}^{equil} \sim 0.215$ correspond to the lower bound $c_s^2 \gtrsim 11/3$ for this model.}
\end{center}
\end{figure}

\begin{figure}[tp] 
\begin{center}$
\begin{array}{c}
\includegraphics[width=0.7\linewidth]{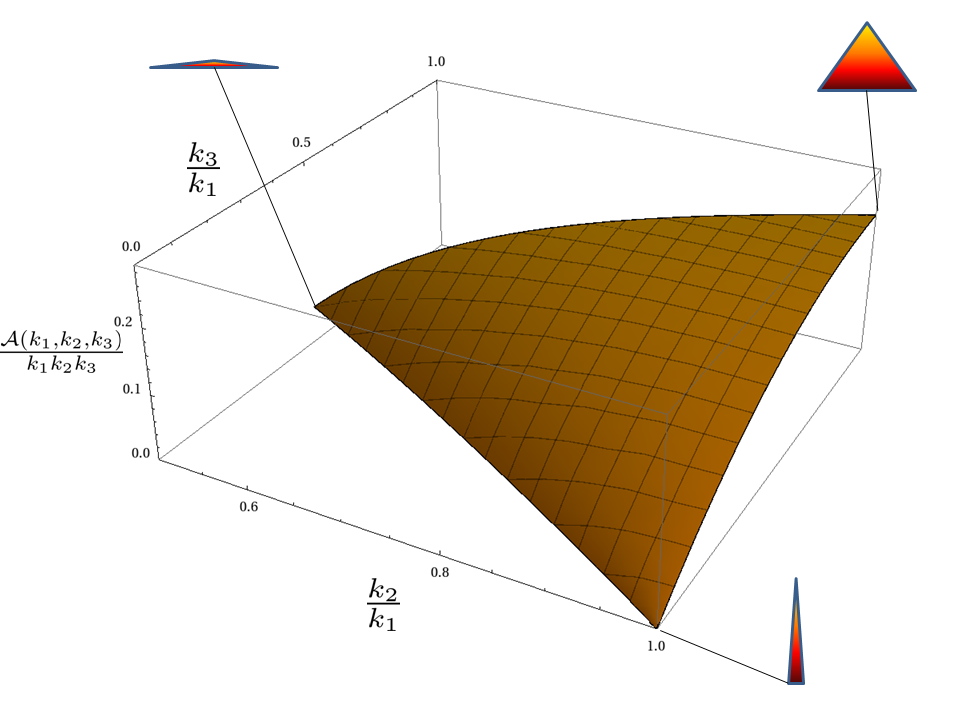} \\
\end{array}$
\end{center}
\caption{
The dimensionless bispectrum ${\cal A}(k_1,k_2,k_3)/k_1 k_2 k_3$ in the slow-roll limit for the exponential potential, $V = V_0 \exp(6\beta\chi^2)$ with $\beta = 1$, corresponding to $c_s \approx 3$ and for $N = 60$. Triangular shapes denote the equilateral, squeezed/local and enfolded limit clockwise from top right. $f_{\rm NL}^{equil} \sim 0.2$ and a predominantly equilateral shape is found. \label{tejo2ampl}}
\end{figure}

The amplitude of the three-point function is controlled by $c_s^2$ and $\lambda/\Sigma$, which are given by
\be
c_s^2 = 1 + 12 \beta \chi^2, \quad\quad\quad \frac{\lambda}{\Sigma} = -\frac{6 \beta^2 \chi^2 (1 + 4 \beta \chi^2)}{c_s^4}.
\ee
Varying $\beta$ and substituting for $c_s^2$ and $\lambda/\Sigma$ $N$ $e$-folds before the end of inflation 
into \eref{eqfnl}, we obtain the dependence of $f_{\rm NL}^{equil}$ on $c_s^2$ which we show in Fig.~\ref{tejo2fnl} for $N = 60$.
In Fig.~\ref{tejo2rfnl} we see how $r$ and $f_{\rm NL}^{equil}$ relate to each other. In particular, we observe that the large values of $c_s^2$ render this model  disfavoured by current bounds on the ratio of tensor to scalar perturbations, $r \lesssim 0.2$ for $N = 60$.\footnote{Note that this bound significantly depends on the assumption of no running spectral index $n_s$. If a running $n_s$ is allowed the bound weakens to $r < 0.49$.} Of course, if we allow for more e-folds of inflation this reduces the relevant $\epsilon$ and hence $r$. A crude estimate for the minimal amount of inflation to bring this model into agreement with current $2\sigma$ bounds on $r$ may be obtained by assuming a speed of sound right at the lower bound of $c_s \gtrsim \sqrt{11/3}$ and $r \sim 0.2$, yielding $N \gtrsim 62$ for the exponential model considered here.
Finally we see in Fig.~\ref{tejo2ampl} that the sign of $f_{\rm NL}^{equil} \sim 0.2$ is positive, albeit with a rather small amplitude. The non-Gaussian shape here is predominantly equilateral, but also picks up contributions in the enfolded/orthogonal limit.\footnote{Note that considering departure from slow-roll could modify this statement, e.g. potentially rendering the shape predominantly enfolded or orthogonal\cite{Noller:2011hd,Noller:2012ed}. For further details on fast-roll corrections also see\cite{Burrage:2011hd,Ribeiro:2012ar}.}

\section{Conclusions}
\label{sec:conclusions}

In this paper, we have considered three-form theories of inflation and their observational footprints. The salient results may be summarised as follows.
\begin{itemize}
\item We have calculated the action for a three-form theory perturbed about a FRW background, isolating the perturbed quadratic and cubic action expressed 
purely in terms of the comoving curvature perturbation $\zeta$, as well as background and three-form quantities. 

\item We then reviewed dualities between various $p$-form models in four dimensional space-times. In particular, these allow the 
linking of the three-form theory under consideration to an effective scalar description with a $P(X,\phi)$ action, due to the fact that the most 
general canonical and minimally coupled 3-form action in 4d has only one independent degree of freedom.
Since the three-form potential is mapped to the scalar field's kinetic term in a non-trivial way, such setups generically result in highly 
non-canonical $P(X,\phi)$ theories, which do, however, all share a simple $\Phi^2$ potential.

\item Despite the complexity of the mapping, we were, however, able to use this duality 
as a formal tool to give an alternative derivation 
for the perturbed quadratic and cubic action. To do so we used existing ``non-Gaussian machinery'' for $P(X,\phi)$ theories, but  
translated the end-result back to three-form quantities (essentially re-expressing $c_s, \Sigma$ and $\lambda$ in terms of the three-form variable $\chi$). 
We showed that the result is identical to that obtained via the direct three-form computation.

\item Finally, using the tools developed, we  explored the observational and particularly non-Gaussian features of two example three-form models: 
\begin{itemize}
\item First, we investigated 
three-form inflation with a power-law potential. This simple 
setup produces a constant speed of sound and generically 
equilateral non-Gaussian shapes with negative $\fnl^{equil}$, similar to 
DBI-type inflation. Observational limits on this form of non-Gaussianity constrain this model. 

\item Secondly, we explored the phenomenology for a three-form inflationary model with an exponential potential. We found 
the non-Gaussianity was once again predominantly of equilateral shape, but in this case was unobservably small. We found, however, 
that the model  can be constrained from limits on the tensor-to-scalar ratio $r$.
\end{itemize}
\item An interesting general result is that the spectral index for all three-form inflationary models is $n_s \approx 0.97$ 
to leading order in slow-roll when $60$ e-folds of inflation occur.  This was shown explicitly for the two models at hand, but will hold much more generally. The value of $n_s$ is therefore uniquely predicted once $N$, the number of e-folds of inflation, is specified, independent of the exact form of the three-form potential and as long as $\epsilon \ll 1$.
\end{itemize}

In conclusion, three-form inflation has a number of interesting features. Simple three-form potentials lead to 
interesting models of inflation with potentially large non-Gaussian signatures, but with a spectral index which is 
within the preferred WMAP bounds. It would be interesting to extend our studies of explicit three-form potentials to more complex
potentials which have no analytic form for the dual scalar theory; to use our framework to compute the trispectrum; to consider a multi three-form action.  We defer these questions to future studies.

\section*{Acknowledgments}
DJM acknowledges support from the Science and Technology Facilities Council
[grant number ST/J001546/1]. JN acknowledges support from the STFC. 
NJN is supported by a Ci\^encia 2008 research contract funded by FCT and  through the projects PEst-OE/FIS/U
I2751/2011 and  grant CERN/FP/116398/2010. The authors would like to thank Antonio de Felice, Tomi Koivisto and Karim Malik for helpful discussions.

\appendix

\section{Appendix I: Dual fields}
\label{appendix1}

Here we give some pedagogical detail on Hodge-dualising $p$-forms in $n$ dimensions. We begin by defining a totally anti-symmetric tensor $\epsilon$ as
\be
  \epsilon^{\alpha_1,\alpha_2...\alpha_p} = \frac{1}{\sqrt{|g|}} \left\{ 
  \begin{array}{l l}
     +1 & \text{if} \; (\alpha_1,\alpha_2...\alpha_p) \; \text{is an even permutation of} \; (1,...d)\\
     -1 & \text{if} \; (\alpha_1,\alpha_2...\alpha_p) \; \text{is an odd permutation of}  \; (1,...d)\\
     0 & \text{otherwise,}\\
  \end{array} \right.
\ee
where $d$ is the number of spatial dimensions  and we define $d_t$ to be the number of temporal dimensions, so $d=3$ and $d_t = 1$ for the cases we will consider in this paper which have signature $(-+++)$. Note that for a diagonal 
metric consequently $(-1)^{d_t} = \text{sign}(g)$, where $g$ is the determinant of $g_{\mu\nu}$.
Lowering all indices with the metric $g_{\mu\nu}$ (note this is a valid procedure since we are explicitly dealing with the tensor $\epsilon$, not its associated tensor density) one finds
\be
  \epsilon_{\alpha_1,\alpha_2...\alpha_p} = (-)^{d_t} \sqrt{|g|} \left\{ 
  \begin{array}{l l}
     +1 & \text{if} \; (\alpha_1,\alpha_2...\alpha_p) \; \text{is an even permutation of} \; (1,...d)\\
     -1 & \text{if} \; (\alpha_1,\alpha_2...\alpha_p) \; \text{is an odd permutation of}  \; (1,...d)\\
     0 & \text{otherwise.}\\
  \end{array} \right.
\ee
A particularly useful identity we use repeatedly is
\be
\epsilon_{\alpha_1....\alpha_n}\epsilon^{\beta_1...\beta_n} = (-)^{d_t} d! \delta^{[\beta_1}_{[\alpha_1} \delta^{\beta_2}_{\alpha_2}...\delta^{\beta_n]}_{\alpha_n]}.
\ee
Let us now consider an arbitrary $p$-form living in an n-dimensional space, where we choose a coordinate basis and write
\be
P \equiv \frac{1}{p!} P_{\alpha_1,...\alpha_p} dx^{\alpha_1} \wedge ...dx^{\alpha_p},
\ee
In terms of the totally antisymmetric tensor $\epsilon$ the Hodge ($\star$) dual of this p-form  is given by
\be
(\star P)_{\alpha_1,...\alpha_{d-p}} = \frac{1}{p!}\epsilon_{\alpha_1,...\alpha_{d-p} \beta_1....\beta_p}P^{\beta_1,...\beta_p}.
\ee
We recall this means any $p$-form has a dual which is a $(d-p)$-form. In particular the three-form $A$ and four-form $F$ that make up~\eqref{ibp} can therefore be expressed as \eref{hodge}.

\section{Appendix II: $\alpha$ dictionary}
\label{appendix2}

In this work we tried as much as possible to use the notation proposed in previous literature. Unfortunately this choice gives rise to an overabundance of variables all named alpha. In this appendix we review their definitions, hoping to clarify their differences.
$\tilde \alpha$ is the scalar perturbation in the lapse function $N$.
$\alpha$ and $\alpha_k$ are respectively the scalar and vector perturbations of the $0ij$ components of the three-form field. $\alpha_0$ is the scalar perturbation in the $ijk$ components of the three-form field. We summarize these definitions in Table \ref{tabela}. 

\begin{table}[h]
\begin{center}
\begin{tabular}{|c|c|}
\hline
$\tilde \alpha$ & $N = 1+ \tilde{\alpha}$ \\
\hline 
$\alpha$ and $\alpha_k$ & $A_{0ij} = a(t)\epsilon_{ijk} (\alpha_{,k}+\alpha_k)$ \\
\hline 
$\alpha_0$ & $A_{ijk} = a(t)^3\epsilon_{ijk}(\chi(t) + \alpha_0)$ \\
\hline 
\end{tabular}
\caption{\label{tabela} Summary of the definitions of $\tilde \alpha$, $\alpha$, $\alpha_k$ and $\alpha_0$.}
\end{center}
\end{table}

\bibliographystyle{JHEP}
\bibliography{paper}

\end{document}